\documentclass[a4paper,fleqn,usenatbib]{mnras}

\usepackage{times}

\usepackage[T1]{fontenc}
\usepackage{ae,aecompl}

\usepackage{url}
\usepackage{graphicx}
\usepackage{amssymb}

\def\H2{{\mathrm{H}_2}}
\def\HI{{\mathrm{H}_{\rm I}}}

\def\SFR{\mathrm{SFR}}
\def\sSFR{\mathrm{sSFR}}

\def\apj{\textit{Astrophys. J.}}                         
\def\araa{\textit{Ann. Rev. Astron. Astrophys.}}  
\def\apj{\textit{Astrophys. J.}}               
\def\apjl{\textit{Astrophys. J. Lett.}}        
\def\apjs{\textit{Astrophys. J. Suppl. Ser.}}  
\def\aap{\textit{Astron. Astrophys.}}          
\def\aapr{\textit{Astron. Astrophys. Rev.}}    
\def\mnras{\textit{Mon. Not. R. Astron. Soc.}} 
\def\nat{\textit{Nature}}              

\title[Massive, quiescent galaxies at cosmic noon]{The formation of massive, quiescent galaxies at cosmic noon}

\author[R. Feldmann et al.]{
Robert Feldmann$^{1}$\thanks{E-mail: feldmann@berkeley.edu (RF)},
Philip F. Hopkins$^{2}$,
Eliot Quataert$^{1}$,\newauthor\phantom{x}
Claude-Andr\'{e} Faucher-Gigu\`{e}re$^{3}$,
and Du\v{s}an Kere\v{s}$^{4}$
\\
$^{1}$Department of Astronomy, University of California, Berkeley, CA 94720-3411, USA\\
$^{2}$TAPIR 350-17, California Institute of Technology, Pasadena, CA 91125, USA\\
$^{3}$Department of Physics and Astronomy and CIERA, Northwestern University, Evanston, IL 60208, USA\\
$^{4}$Center for Astrophysics and Space Sciences, University of California, San Diego, CA 92093, USA
}

\date{Accepted XXX. Received YYY; in original form ZZZ}

\pubyear{2016}

\begin{document}
\label{firstpage}
\pagerange{\pageref{firstpage}--\pageref{lastpage}}
\maketitle

\begin{abstract}
The cosmic noon ($z\sim{}1.5-3$) marked a period of vigorous star formation for most galaxies. However, about a third of the more massive galaxies at those times were quiescent in the sense that their observed stellar populations are inconsistent with rapid star formation. The reduced star formation activity is often attributed to gaseous outflows driven by feedback from supermassive black holes, but the impact of black hole feedback on galaxies in the young Universe is not yet definitively established. We analyze the origin of quiescent galaxies with the help of ultra-high resolution, cosmological simulations that include feedback from stars but do not model the uncertain consequences of black hole feedback. We show that dark matter halos with specific accretion rates below $\sim{}0.25-0.4$ Gyr$^{-1}$ preferentially host galaxies with reduced star formation rates and red broad-band colors. The fraction of such halos in large dark matter only simulations matches the observed fraction of massive quiescent galaxies ($\sim{}10^{10}-10^{11}$ $M_\odot$). This strongly suggests that halo accretion rate is the key parameter determining which massive galaxies at $z\sim{}1.5-3$ become quiescent. Empirical models that connect galaxy and halo evolution, such as halo occupation distribution or abundance matching models, assume a tight link between galaxy properties and the masses of their parent halos. These models will benefit from adding the specific accretion rate of halos as a second model parameter.
\end{abstract}



\section{Introduction}

There is mounting evidence that star formation in galaxies is tied to the accretion of gas from intergalactic distances \citep[e.g.,][]{2009Natur.457..451D, 2010Natur.467..811C, 2013ApJ...772..119L, 2014A&ARv..22...71S,2014ApJ...786..106M, 2015arXiv150804842R, 2015ApJ...799...13B, 2015Natur.525..496N}. Hence, reduced gas accretion onto galaxies could potentially be responsible for the reduced star formation rates (SFRs) of quiescent galaxies \citep{2015MNRAS.446.1939F}. The reduced supply of gas to galaxies and halos would also make it easier for additional processes, e.g., black hole feedback 
(e.g., \citealt{1997ApJ...487L.105C, 2005Natur.433..604D, 2006MNRAS.366..397S, 2011MNRAS.414..195T, 2013ARA&A..51..511K}), to fully suppress any remaining star formation activity in quiescent galaxies \citep{2006MNRAS.368....2D, 2006MNRAS.370.1651C}. Numerical simulations are the tool of choice to test this proposed picture, but until now no existing simulation produced both a galaxy sample of the necessary size for statistical analysis \emph{and} properly resolved and modeled the relevant physical processes that take place in galaxies at $z\sim{}1.5-3$. In addition, cosmological, hydrodynamical simulations have a long history of struggling to reproduce key properties of observed galaxies such as their typical stellar masses and star formation rates \citep{2012MNRAS.423.1726S}. The present study, MassiveFIRE, overcomes these challenges by adopting the accurate physical modeling of the \emph{Feedback In Realistic Environments} (FIRE) project \citep{2014MNRAS.445..581H} and by applying it, for the first time, to a population of massive galaxies.

We have simulated the evolution of 35 massive galaxies for the first 4 billion years after the Big Bang (until redshift $z\geq{}1.67$).
The galaxy sample is extracted from 17 distinct sub-regions, each containing at least one dark matter (DM) halo with a mass in the range $3\times{}10^{12}-3\times{}10^{13}$ $M_\odot$, embedded in a representative volume of the Universe. The sub-regions sample the full range of cosmological assembly histories of halos harboring massive galaxies, i.e., galaxies with stellar masses larger than about $10^{10}$ $M_\odot$. Our simulations are run with the hydrodynamics and gravity solver GIZMO \citep{2015MNRAS.450...53H} at ultra-high spatial ($\sim{}10$ pc) and mass resolution ($m_{\rm gas}=3.3\times{}10^4$ $M_\odot$ at high resolution, $2.7\times{}10^5$ $M_\odot$ at medium resolution) in P-SPH mode. The high numerical resolution allows us to model reliably many of the relevant processes that take place in the interstellar medium of galaxies. Stellar feedback processes such as energy and momentum injection from supernovae, stellar winds, photo-heating, and radiation pressure interact in a non-linear manner \citep{2012MNRAS.421.3522H} and are all included in our simulations with little reliance on tunable parameters. Feedback from supermassive black holes (SMBH) is not included. We will present the setup and methodology of our simulations in more detail in Feldmann et al. in prep.

We showed in previous work that the computational approach of this paper reproduces well the integral properties of lower mass galaxies ($M*\lesssim{}3\times{}10^{10}$ $M_\odot$) since cosmic noon. For instance, we reported on the SHMR \citep{2014MNRAS.445..581H}, on the stellar mass -- metallicity relation \citep{2015arXiv150402097M}, and on the properties of galactic outflows driven by stellar feedback \citep{2015MNRAS.454.2691M} finding good agreement with available observations. We also showed that the $H_{\rm I}$ covering fractions in $\lesssim{}10^{12}$ $M_\odot$ halos at cosmic noon match observations \citep{2015MNRAS.449..987F}. We will report corresponding properties of MassiveFIRE galaxies in upcoming work.

\vspace{-0.5cm}
\section{Results and Discussion}

\begin{figure} 
\begin{tabular}{c} 
\includegraphics[width=78mm]{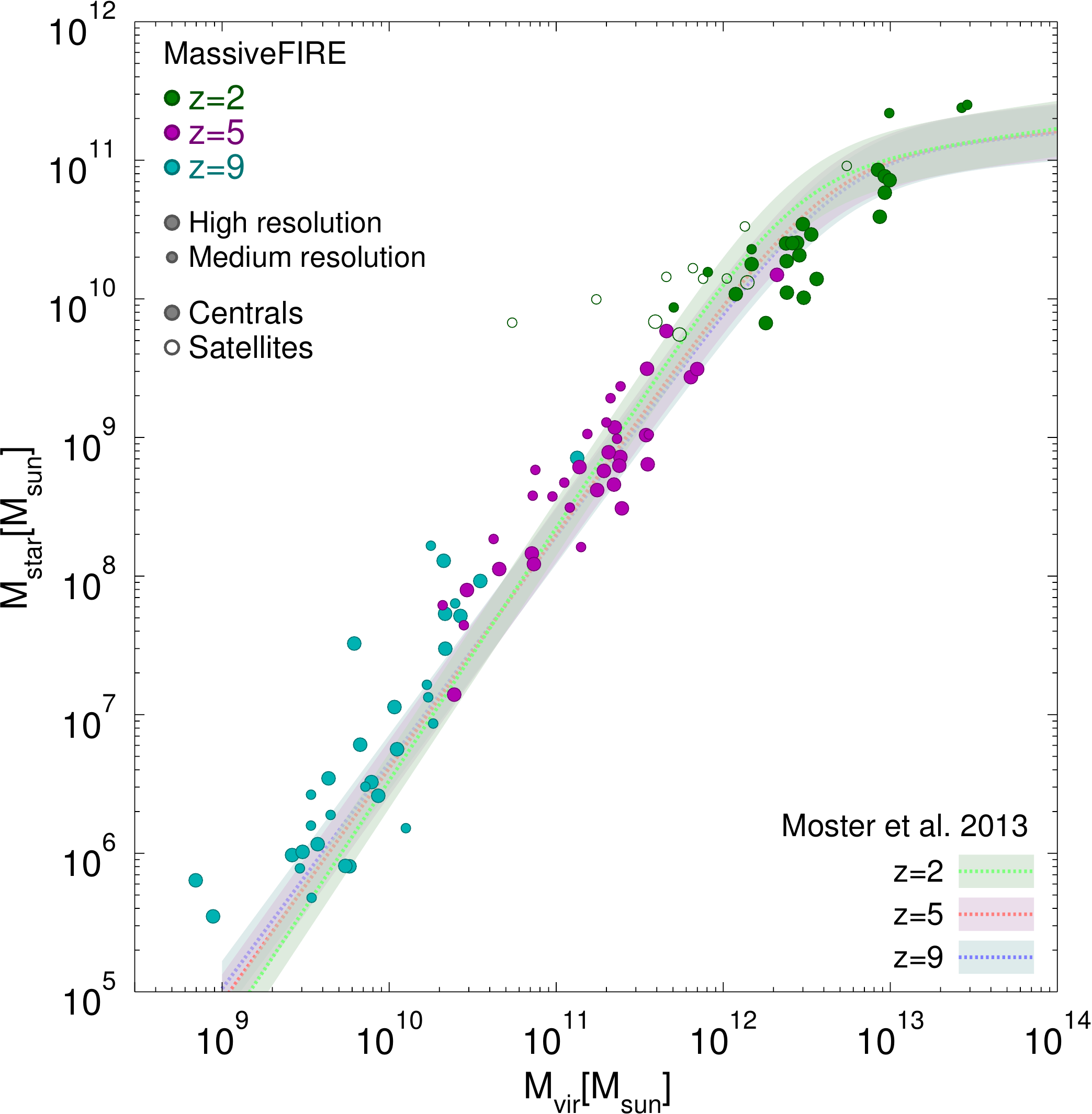}
\end{tabular}
\caption{Stellar-to-halo-mass relation (SHMR) of MassiveFIRE galaxies at $z=2$ (green circles), $z=5$ (purple circles), and $z=9$ (cyan circles). Central (satellite) galaxies at each redshift are shown by filled (empty) circles, and large (small) circles denote galaxies simulated at high (medium) numerical resolution. Dotted lines show an empirical estimate \protect\citep{2013MNRAS.428.3121M} of the SHMR and its extrapolation to high redshifts and low stellar masses. The  $1-\sigma$ scatter of individual galaxies above and below the mean relation is about 0.2 dex \protect\citep{2013ApJ...771...30R} (shaded region). Satellite galaxies tend to lie to the left of the relation as their DM halos are often tidally stripped. MassiveFIRE galaxies have stellar masses in fair agreement with the empirically derived SHMR.} 
\label{fig:Mhalo_Mstar}
\end{figure}

Figure \ref{fig:Mhalo_Mstar} compares the stellar masses of MassiveFIRE galaxies with those of actual galaxies that reside in DM halos of  similar mass. Stellar masses of MassiveFIRE galaxies agree, to within a factor of $\sim{}2$, with the empirically inferred estimate of the stellar-to-halo-mass relation (SHMR) \citep{2013MNRAS.428.3121M} at $z=2$ and its extrapolation to higher redshifts. We note, however, that the exact functional form of the SHMR relation differs somewhat among studies \citep{2013MNRAS.428.3121M, 2013ApJ...770...57B, 2014MNRAS.438.2578G}.

\begin{figure}
\begin{tabular}{c}
\includegraphics[width=78mm]{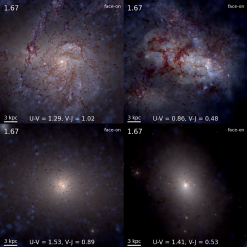}
\end{tabular}
\caption{Color-composite images of four central galaxies in the MassiveFIRE sample as they would appear in rest-frame U, V, and J bands $\sim{}$4 billion years after the Big Bang ($z=1.67$). Galaxies are shown face-on with each image spanning 30 kpc on each side. (Top left) a star forming, disk galaxy, (top right) a star forming, irregular galaxy, (bottom row) two examples of quiescent, early type galaxies. Star forming and quiescent galaxies differ in their colors, morphologies, and levels of dust extinction.}
\label{fig:QSFexamples}
\end{figure}

Fig.~\ref{fig:QSFexamples} shows four typical example galaxies from MassiveFIRE. Each panel displays a composite image (in rest-frame U, V, and J broad-band filters) of the dust reprocessed star light. Following conventional practice \citep{2007ApJ...655...51W, 2009ApJ...691.1879W, 2011ApJ...735...86W} we classify galaxies as quiescent based on their U-V and V-J broad band colors; specifically, if $U-V > 1.2$, $V-J < 1.4$, and $U-V > 0.88\times{}(V-J)+0.59$. Our sample contains 8 quiescent and 15 star forming central galaxies (galaxies that dominate the central potential well of their host DM halo), and 4 quiescent and 8 star forming satellites (galaxies orbiting a central galaxy).
Star forming galaxies have younger stellar populations than quiescent galaxies, resulting in bluer intrinsic colors, although interspersed dust lanes extinct and redden the light along particular lines of sight (see top panels of Fig.~\ref{fig:QSFexamples}). Fortunately, the color-color classification is relatively insensitive to the amount of dust reddening. Most star forming galaxies in our sample have a late type morphology with either large stellar and gas disks (half mass radii $>3$ kpc) or irregular shapes, in agreement with observations \citep{2013ApJ...774...47L}. In contrast, quiescent galaxies often have an early type morphology with a more compact stellar distribution \citep{2014ApJ...788...28V} and contain only low levels of dust and cold gas (bottom row of Fig.~\ref{fig:QSFexamples}). 

\begin{figure} 
\begin{tabular}{c}
\includegraphics[width=80mm]{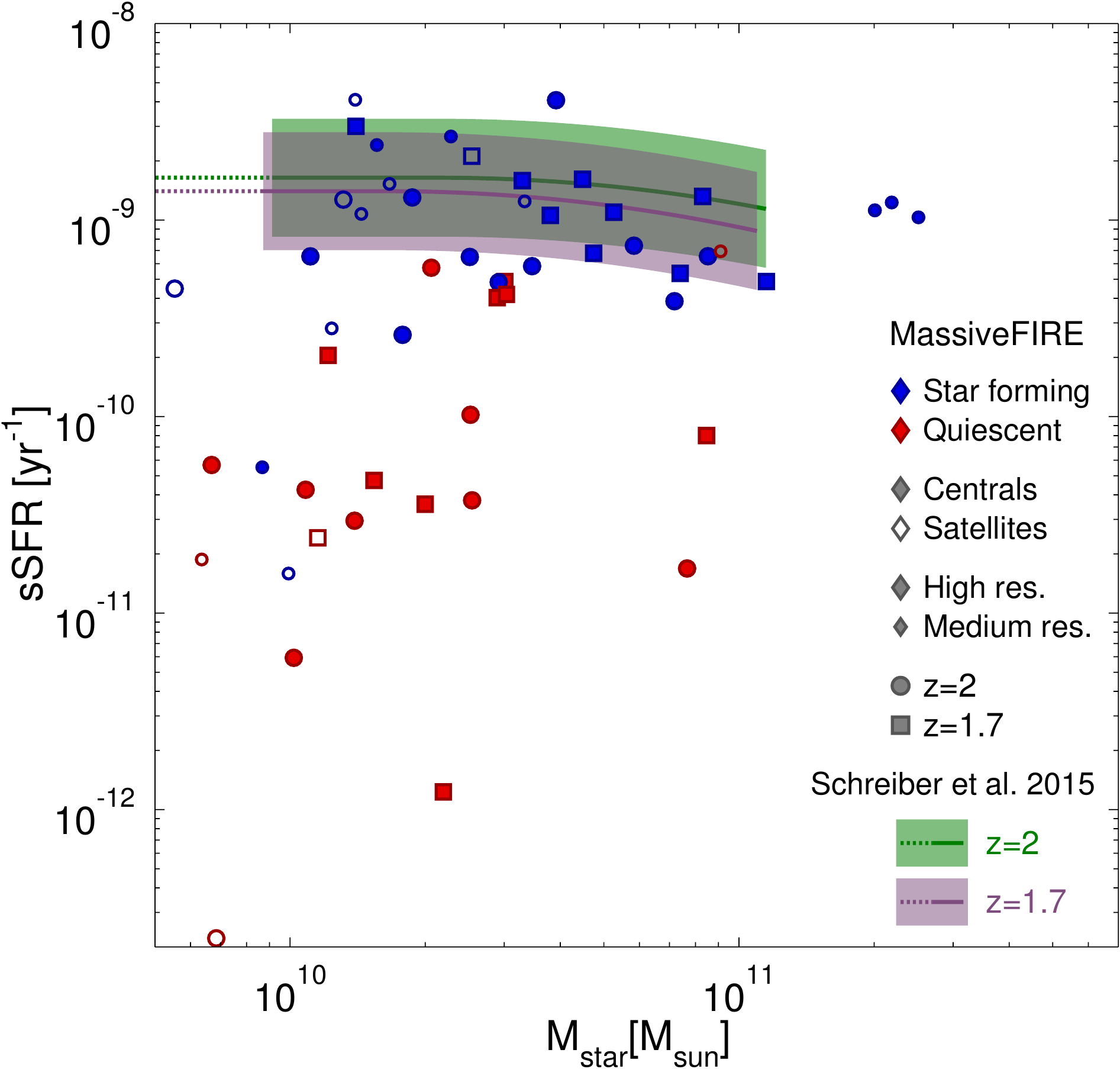}
\end{tabular}
\caption{Specific SFR within the central 5 kpc of MassiveFIRE galaxies as function of stellar mass. SFRs are averaged over the past 100 Myr. Star forming and quiescent galaxies  are shown by blue and red symbols respectively (the classification is based on rest-frame U, V, and J broad band fluxes). Lines denote the location of the star forming sequence inferred from rest-frame ultraviolet and infrared observations \protect\citep{2015A&A...575A..74S}. The $1-\sigma$ scatter of individual galaxies above and below the star forming sequence at $z\sim{}2$ is about 0.3 dex \protect\citep{2007ApJ...670..156D, 2012ApJ...754L..29W, 2015A&A...575A..74S} (shaded region). MassiveFIRE galaxies classified as star forming have specific SFRs consistent with the observed star forming sequence. Star formation in quiescent galaxies, however, proceeds at much lower rates than in star forming galaxies of comparable stellar mass.}
\label{fig:Mstar_sSFR}
\end{figure}

The specific SFR for both star forming and quiescent galaxies in the MassiveFIRE sample is shown in Figure \ref{fig:Mstar_sSFR}. We measure the specific SFR in 5 kpc radii to roughly mimic aperture based flux measurements \citep{2011ApJ...735...86W, 2015A&A...575A..74S} and to minimize potential contributions from low mass satellite galaxies. Specific SFRs change typically by less than 0.1 dex if measured within a radius of 0.1 $R_{\rm vir}$ instead.
Star forming galaxies at cosmic noon have high specific SFRs of the order of $\sim{}1$ Gyr$^{-1}$, while quiescent galaxies form stars at significantly lower specific rates. We note that in most cases the SFRs of galaxies classified as quiescent remain low for extended periods of time ($>3\times{}10^8$ yr). The specific SFRs of MassiveFIRE galaxies are in good agreement with observations \citep{2015A&A...575A..74S, 2011ApJ...739...24B}. 

\begin{figure} 
\begin{tabular}{cc} 
\includegraphics[width=78mm]{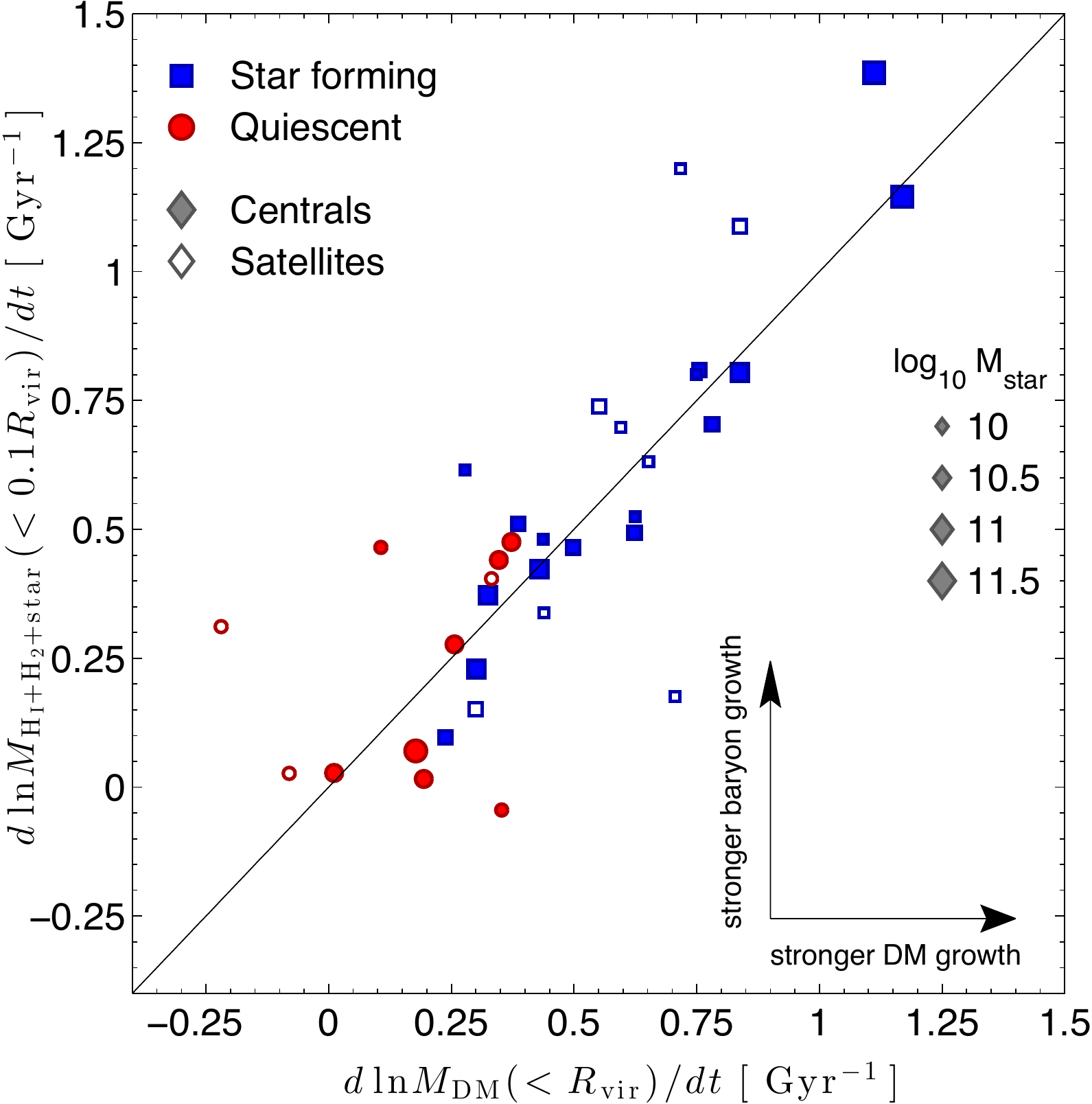} 
\end{tabular}
\caption{Comparison between the growth rate of baryonic masses (stars, $\HI$, and $\H2$) of galaxies and the DM masses of their parent halos. Red circles and blue squares show quiescent and star forming galaxies in MassiveFIRE, respectively. The classification is based on rest-frame U-V and V-J colors \protect\citep{2011ApJ...735...86W} appropriate for $z\sim{}2$. Filled and empty symbols denote galaxies that are centrals or satellites by the final snapshot of the simulation ($z=1.7-2$). Symbol sizes reflect stellar masses. For central galaxies, growth rates and colors are computed at the final snapshot of each simulation. For galaxies that become satellites by $z\sim{}2$, we compute growth rates and colors in the last snapshot before they enter their host halo. The solid line marks a 1:1 relationship and is not a fit. Galaxies residing at the centers of fast growing halos ($d\ln{}M_{\rm DM}/dt\gtrsim{}0.4$ Gyr$^{-1}$) are essentially always strongly star forming. In contrast, slowly growing (or even shrinking) halos typically harbor quiescent galaxies.}
\label{fig:accfit_MDM_Mcbar}
\end{figure}

We fit the growth history of the (cold) baryonic mass ($M_{\rm bar} = M_\HI + M_\H2 + M_*$) of each galaxy with a modified exponential $\propto{}(1+z)^{\beta}e^{-\gamma{}z}$ over
an extended redshift range starting from $z=7$ down to either the final simulation snapshot or to the last snapshot at which the galaxy is still a central, whichever comes first. Gas and stars within $10\%$ of the virial radius from the center of a galaxy are considered part of that galaxy. Our results do not change qualitatively if we use a 50\% larger or smaller radius instead. We similarly fit the growth of the DM mass, $M_{\rm DM}$, contained within the virial radius of the halos surrounding these galaxies. In Fig.~\ref{fig:accfit_MDM_Mcbar} we plot $d\ln{}M_{\rm bar}/dt$ and $d\ln{}M_{\rm DM}/dt$ for both star forming and quiescent galaxies in MassiveFIRE, linking the growth of DM halos to the growth of galaxies residing at the centers of those halos. The figure demonstrates that most galaxies at cosmic noon grow on the same timescale as the DM halos they live in, complementing previous work that showed that baryonic masses and DM masses of \emph{halos} assemble on similar timescales \citep{2011MNRAS.417.2982F}. This is a non-trivial result as galaxies contain only a small fraction, less than a fifth, of the baryons in halos \citep{2012ApJ...759..138P}.

At cosmic noon, galaxies with declining SFRs are on their way to becoming quiescent. Hence, we may introduce an alternative definition of ``quiescence'' that is not based on broad-band colors, but on the star formation history of galaxies. In particular, by manipulating the standard equations for one-zone galaxy models including inflow, outflow, star formation, and gas build up (e.g., \citealt{2013ApJ...772..119L, 2015MNRAS.449.3274F}) the condition of a declining SFR can be shown to be equivalent to $d\ln{}M_{\rm bar}/dt <\mathcal{X}_{\rm crit}\equiv[1-R+dt_{\rm dep}/dt]/[t_{\rm dep} + \sSFR^{-1}]$. Here, $t_{\rm dep}=(M_\HI+M_\H2)/\SFR$ is the gas depletion time, $R$ is the return fraction of gas from evolved stellar populations, and $\sSFR(t)=\SFR/M_*=A\,\sSFR_{\rm MS}(M_*(t),t)$ is the specific SFR. $\sSFR_{\rm MS}$ is the specific SFR of galaxies of the same mass on the star forming sequence and $A$ can be derived from the criticality condition $d{}\SFR/dt=0$.
Upon inserting values appropriate for galaxies in the $M_*\sim{}10^{10}-10^{11}\,M_\odot$ range, we find that such galaxies should be reducing their star formation activity, and thus becoming quiescent, when $d\ln{}M_{\rm bar}/dt \lesssim{} 0.25-0.4$ Gyr$^{-1}$.

In agreement with this analysis, $d\ln{}M_{\rm bar}/dt \sim{} 0.4$ Gyr$^{-1}$ largely separates star forming from quiescent galaxies in the MassiveFIRE sample, as shown in Fig.~\ref{fig:accfit_MDM_Mcbar}. As galaxies and halos grow on very similar timescales (see Fig. \ref{fig:accfit_MDM_Mcbar}), we can re-interpret this result in terms of the specific growth rates of DM halos, i.e., $d\ln{}M_{\rm DM}/dt\lesssim{}0.4$ Gyr$^{-1}$ is a necessary condition for a halo to host a quiescent galaxy at its center at cosmic noon. We propose that the majority of moderately massive, quiescent galaxies in the young Universe form via this mechanism, i.e., they reside in the sub-set of halos that accrete gas from the cosmic web at such low rates that they cannot maintain SFRs characteristic of typical star forming galaxies \citep{2015A&A...575A..74S}. As discussed more below, numerous halos undergoing such ``cosmological starvation'' \citep{2015MNRAS.446.1939F} should exist given the variations in the gravity-driven collapse histories of DM halos \citep{2009MNRAS.398.1858M}.

\begin{figure}
\begin{tabular}{c}
\includegraphics[width=79mm]{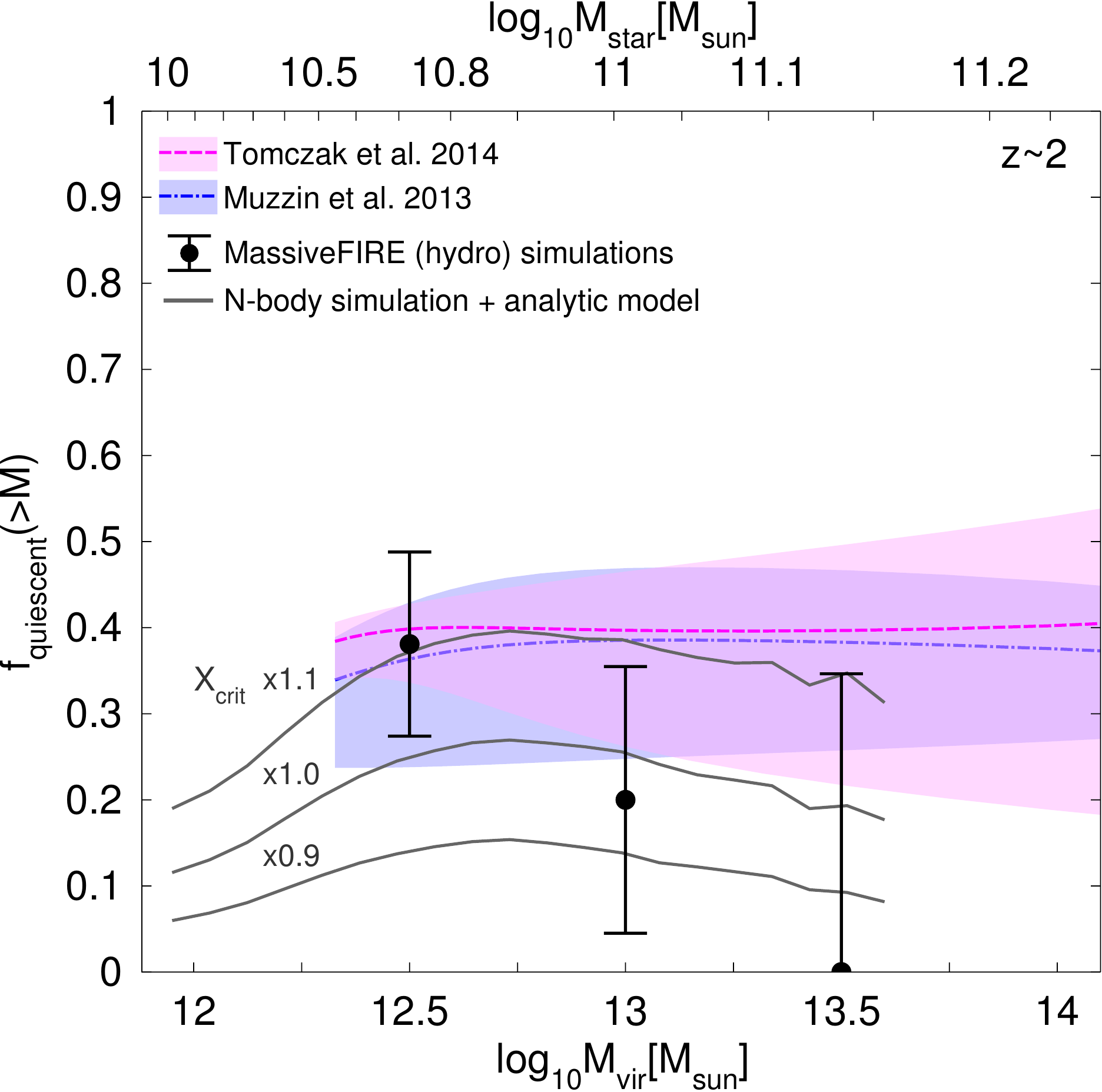}
\end{tabular}
\caption{Fraction of quiescent, central galaxies residing in halos above a given mass. The fractions predicted by MassiveFIRE are shown by filled circles. Error bars indicate $1-\sigma$ standard deviations based on a binomial distribution with the same sample size and quiescent fraction as in MassiveFIRE. For the largest mass bin we assume a 40\% quiescent fraction to compute the error bar. Our simulations predict that about a third of massive galaxies at cosmic noon are quiescent. Solid lines show the fraction of halos with specific growth rates below the critical value required for quiescent galaxies (from Fig. \ref{fig:accfit_MDM_Mcbar}), $0.9-1.1\times\mathcal{X}_{\rm crit}\sim{}0.25-0.4$ Gyr$^{-1}$, based on the Millennium $N$-body simulation \protect\citep{2005Natur.435..629S, 2009MNRAS.398.1858M}. These theoretical estimates agree reasonably well with the observed quiescent fraction derived from stellar mass functions of quiescent and star forming galaxies over the $z=1.5-2.5$ range (dashed \protect\citep{2014ApJ...783...85T} and dot-dashed \protect\citep{2013ApJ...777...18M} lines and shaded regions).} 
\label{fig:quiescent_fraction}
\end{figure}

$35\%$ of the central galaxies and 34\% of all galaxies in our sample are quiescent. These numbers compare favorably with observations of quiescent fractions of $25-50\%$ over a broad stellar mass range \citep{2014ApJ...783...85T,2013ApJ...777...18M}; see Fig.~\ref{fig:quiescent_fraction}. We note that the cumulative fraction of quiescent galaxies in MassiveFIRE is lower at larger stellar masses, while the observed fraction remains remains relatively flat. This difference could point towards missing physics in our simulations, such as black hole feedback, or it could be an artifact related to the low number of galaxies (two) in our highest halo mass bin. Hence, whether cosmological starvation is an effective quenching mechanisms for galaxies residing in the most massive halos at $z=2$ ($M_{\rm halo}>10^{13}$ $M_\odot$, $n<10^{-5}$ Mpc$^{-3}$) remains to be studied in future work.

Fig.~\ref{fig:quiescent_fraction} also plots the fraction of halos with low specific growth rates based on a large-volume cosmological $N$-body simulation \citep{2005Natur.435..629S, 2009MNRAS.398.1858M}. The fraction of halos with $d\ln{}M_{\rm DM}/dt<0.9-1.1\,\mathcal{X}_{\rm crit}$ matches fairly well the observed quiescent fraction. The former declines slightly towards the largest stellar masses, indicating that additional physics besides cosmological starvation is likely involved in shutting down star formation in the most massive galaxies ($M_*>10^{11}$ $M_\odot$). Fig.~\ref{fig:accfit_MDM_Mcbar} shows that there is overlap between star forming and quiescent galaxies at intermediate specific growth rates. We find that the fraction of slowly accreting halos still matches the observed fraction of quiescent galaxies even if a sizable fraction of such halos, e.g., a third, host star forming galaxies.

\vspace{-0.6cm}
\section{Conclusions}

The star formation activity of massive galaxies in a young Universe is ultimately fueled by the accretion of intergalactic gas \citep{2005MNRAS.363....2K, 2009Natur.457..451D, 2010MNRAS.404.1355D, 2013MNRAS.429.3353N}. By limiting the supply of gas to galaxies and halos, cosmological starvation makes it much easier for additional processes, e.g., feedback from black holes, to fully counteract hot gas cooling and to heat or eject any remaining cool gas. Cosmological starvation thus enables the formation of quiescent galaxies with red broad-band colors and reduced SFRs at cosmic noon. However, it may be a necessary but not a sufficient condition for \emph{completely} shutting-down star formation in such galaxies.

The different accretion histories of quiescent and star forming galaxies in halos of the same mass have a number of observational consequences. First, as quiescent galaxies are assembled earlier, they will be surrounded by more evolved satellite populations. In particular, orbital decay and tidal stripping \citep{2005ApJ...624..505Z} should reduce the number of satellites of a given stellar mass, and the longer exposure \citep{2011ApJ...736...88F} to the hot atmospheres of massive galaxies may explain the increased fraction of satellites with low SFRs \citep{2006MNRAS.366....2W}.
Second, we expect that the dominant halos of over-dense environments should have large accretion rates and, thus, should host vigorously star forming galaxies. In contrast, quiescent galaxies at those redshifts should preferentially reside in average or below average environments.
This idea is corroborated by our finding that 50\% (25\%) of the quiescent central galaxies vs 13\% (73\%) of the star forming central galaxies in our sample reside in the lower (upper) quartile of the local environment density.
Third, the clustering of halos of a given mass depends on their formation time, the so-called assembly bias \citep{2006ApJ...652...71W}. As massive DM halos are less clustered if they collapsed earlier, we predict that massive, quiescent galaxies at cosmic noon have a \emph{lower} clustering amplitude than star forming galaxies residing within halos of the same mass. 
Finally, we speculate that age-matching \citep{2013MNRAS.435.1313H}, an empirical correlation between galaxy colors and halo formation time, has its physical origin in cosmological starvation.

\vspace{-0.6cm}
\section*{Acknowledgements}
RF was supported in part by NASA through Hubble Fellowship grant HF2-51304.001-A awarded by the Space Telescope Science Institute, which is operated by the Association of Universities for Research in Astronomy, Inc., for NASA, under contract NAS 5-26555, in part by the Theoretical Astrophysics Center at UC Berkeley, and by NASA ATP grant 12-ATP-120183. Support for PFH was provided by an Alfred P. Sloan Research Fellowship, NASA ATP Grant NNX14AH35G, and NSF Collaborative Research Grant \#1411920 and CAREER grant \#1455342. CAFG was supported by NSF through grants AST-1412836 and AST-1517491, by NASA through grant NNX15AB22G, and by Northwestern University funds. DK was supported by NSF grant AST-1412153. EQ was supported by NASA ATP grant 12-ATP-120183, a Simons Investigator award from the Simons Foundation, and the David and Lucile Packard Foundation. Simulations were run with resources provided by the NASA High-End Computing (HEC) Program through the NASA Advanced Supercomputing (NAS) Division at Ames Research Center, proposal SMD-14-5492. Additional computing support was provided by HEC allocation SMD-14-5189 and NSF XSEDE allocation TG-AST120025.
\vspace{-0.6cm}


\bibliographystyle{mnras}

\bsp	
\label{lastpage}

\end{document}